\documentclass[aoas,MSNbibl,nameyear,dvips]{arximspdf}
\usepackage{dcolumn}
\usepackage{graphicx}
\usepackage{url,breakurl}

\doi{10.1214/14-AOAS753} 
\volume{8}
\issue{3}
\pubyear{2014}
\firstpage{1372}
\lastpage{1394}
\docsubty{FLA}

\makeatletter
\newcolumntype{d}[1]{D{.}{.}{#1}}

\makeatother

\begin{document}
\begin{frontmatter}

\title{Change points and temporal dependence in reconstructions of
annual temperature:\\ Did Europe experience a Little Ice Age?}
\runtitle{European Little Ice Age}

\begin{aug}
\author[A]{\fnms{Morgan}~\snm{Kelly}\corref{}\ead[label=e1]{morgan.kelly@ucd.ie}}
\and
\author[A]{\fnms{Cormac}~\snm{\'O Gr\'ada}\ead[label=e2]{cormac.ograda@ucd.ie}}
\runauthor{M. Kelly and C. \'O Gr\'ada}
\affiliation{University College Dublin}
\address[A]{School of Economics\\
University College Dublin\\
Belfield, Dublin 4\\
Ireland\\
\printead{e1}\\
\phantom{E-mail: }\printead*{e2}} 
\end{aug}

\received{\smonth{12} \syear{2011}}
\revised{\smonth{5} \syear{2014}}

%
\begin{abstract}
We analyze the timing and extent of Northern European temperature
falls during the Little Ice Age, using standard temperature reconstructions.
However, we can find little evidence of temporal dependence or structural
breaks in European weather before the twentieth century. Instead,
European weather between the fifteenth and nineteenth centuries resembles
uncorrelated draws from a distribution with a constant mean (although
there are occasional decades of markedly lower summer temperature)
and variance, with the same behavior holding more tentatively back
to the twelfth century. Our results suggest that observed conditions
during the Little Ice Age in Northern Europe are consistent with random
climate variability. The existing consensus about apparent cold conditions
may stem in part from a Slutsky effect, where smoothing data gives
the spurious appearance of irregular oscillations when the underlying
time series is white noise.
\end{abstract}

%
\end{frontmatter}

\section{Introduction}\label{sec1}

The Little Ice Age is generally seen as a major event in European
climatic history, causing Swiss glaciers to advance, the Thames in
London to freeze, grape growing to disappear from England and wheat
from Norway, and the Norse colonies in Greenland to perish. The goal
of this paper is to estimate the magnitude and timing of climatic
deteriorations during the Little Ice Age in Europe by using a variety
of standard summer and winter temperature reconstructions: for
Central Europe since 1500 [\citet{Dobrovolny2009}]; the Netherlands from 1301
[\citet{vanE01}]; Switzerland from 1525 [\citet{cPfi92}]; and England
from 1660 [\citet{gMan74}]. However, contrary to the existing consensus
of a European Little Ice Age, we can find little evidence for change
points or temporal dependence in the weather series that we examine.

Starting with standard classical tests for change points in mean
tem\-perature---\citet{jBpP98}
breakpoints and \citet{eVaO07} binary segmentation, which we show
can detect changes of one standard deviation lasting a generation,
and 0.5 standard deviations lasting a century---we find that all winter
series break around 1900, as does Central European summer temperature.
However, there is no indication of any shift in mean temperature,
apart from a brief fall in Switzerland during the 1810s, prior to
this. Using the Bayesian change point analysis of \citet
{BarryHartigan1993},
which has greater ability to detect short deviations, we find that
while winters before 1900 are stable, there are occasional decades
of markedly reduced summer temperature (the 1590s in Central Europe,
the 1690s in England and the 1810s everywhere), but again no evidence
of sustained falls in mean temperature.

However, temperature changes during a European Little Ice Age are
more likely to have been gradual trends than discrete jumps. Therefore,
the core of this paper consists of tests for temporal dependence;
looking specifically at whether annual temperature in Europe displays
conditional mean independence or martingale differences, given the
past history of a stationary series $ \{ Y_{t} \} $ with
expectation $\mu$, the best forecast of its current value is its
unconditional mean: \mbox{$E (Y_{t}|Y_{t-1},Y_{t-2},\ldots)=\mu$}.

Developed in financial econometrics to test the Efficient Markets
Hypothesis that changes in asset prices should be unpredictable, there
are three principal categories of these tests: Portmanteau tests that
look at the sum of autoregressive coefficients in the data; variance
ratio tests that look at how fast the variance of a series grows as
the number of observations rises; and spectral tests that look for
departures from a straight line spectrum. While little known outside
econometrics, these tests have considerable power to detect a wide
variety of patterns of temporal dependence. We apply recent versions
of the three types of test to European temperature series and find
few departures from conditional mean independence, with these departures
driven by pre-1700 reconstructions.

In summary, then, annual temperature reconstructions for Northern
Europe do not appear to exhibit temporal dependence or structural
breaks consistent with the occurrence of a European Little Ice Age.
To avoid confusion about this result, it is useful to bear in mind
that, in contrast to the hemispheric or global scale on which climatologists
usually operate, the scale of our analysis is regional. A fundamental
lesson of efforts over the past twenty years to construct estimates
of global temperature is that conditions in individual regions are
extremely noisy signals of global conditions. The fact that European
weather has been fairly constant between the late Middle Ages and
late nineteenth century no more implies that wider Northern Hemisphere
temperatures were constant during this period than the absence of
marked rises in European summer temperature during the twentieth century
implies that global warming was not occurring. Similarly, the slowly
changing nature of forcing variables that drive global climate (solar
output, Earth orbit, land cover, greenhouse gas concentrations; with
more rapid movements associated with volcanic activity) would lead
us to expect that annual Northern Hemisphere temperatures display
considerable autocorrelation, but this is not incompatible with the
absence of autocorrelation that we find in Europe.%
\footnote{The autocorrelation for the Climatic Research Unit's Northern
Hemisphere series \citet{entry-1} from 1850 to 1949 is 0.4 for July with $p$-value
of $p=0.001$, but $-$0.1 for January with $p=0.2$.}

That our findings run counter to the existing consensus of a European
Little Ice Age may reflect the fact that our analysis is based on
unsmoothed data. This is in contrast to the common practice in climatology
of smoothing data using a moving average or other filter to extract
long run climate signals from noisy local weather observations. When
data are uncorrelated, as the annual European weather series we examine
appear to be, such smoothing can introduce the appearance of irregular
oscillations: a Slutsky effect. For an example of this in climatology
see Figure~2 in \citet{mMan02} where the fall in smoothed Central
England temperature ``which best defines the European Little Ice
Age'' is explicitly highlighted. We illustrate the Slutsky effect
in Figure~\ref{figlowcountries} below for the four temperature
series: while annual temperatures are uncorrelated, when smoothed
with a 30 year moving average the data appear to show marked episodes
of lower temperature consistent with a European Little Ice Age.

The rest of the paper is as follows. The traditional view of the European
Little Ice Age is outlined in Section~\ref{secThe-Little-Ice}, along
with descriptions of the data sources. Section~\ref{secChange-Points-in}
applies classical and Bayesian change point tests to the temperature
reconstructions and finds little evidence of sustained changes before
the twentieth century, although there are some decades of markedly
lower summer temperature. Section~\ref{secConditional-Independence}
looks for temporal dependence in annual temperature using tests of
conditional mean independence, while Section~\ref{secThe-Slutsky-Effect}
outlines the Slutsky effect.

Supplement~A [\citet{KOG14a}], available online, finds similar behavior
in other weather records: instrumental records from European cities
since 1700; the reconstruction of average temperature across all of
Europe since 1500 by \citet{LDX04}; German weather since AD 1000
[\citet{GlaRie09}]; and English and Swiss precipitation. It also tests
for the presence of changing volatility in the annual reconstructions
examined here by testing for autoregressive conditional heteroskedasticity.

\section{The Little Ice Age}\label{secThe-Little-Ice}

Originally coined in by \citet{fMat39} to describe the increased
extent of glaciers over the last 4000 years, the term ``Little
Ice Age'' now usually refers instead to a climatic
shift toward colder weather occurring during the second millennium.
There is a consensus among climatologists that much of the Northern
Hemisphere above the tropics experienced several centuries of reduced
mean summer temperatures, although there is some variation over dates,
with \citet{mMan02} suggesting the period between the fifteenth and
nineteenth centuries, \citet{kBjM05} 1570--1900, and \citet{MEM+09}
between 1400 and 1700.

\citet{hWan08} identify a fall in global temperatures between around
1350 and 1850 associated with variations in the Earth's orbit, lowered
solar output and several large volcanic eruptions. More recently,
the \citet{PAGES13} finds that timing of peak cold periods in the
last millennium varies regionally, with Europe [where reconstructions
are based on mountain and Scandinavian tree rings, and the \citet
{Dobrovolny2009}
documentary reconstruction that we analyze here] showing a transition
to a cooler climate around 1250.\footnote{If the European Little Ice Age occurred as a sudden and
permanent jump
to worse conditions, these start points precede most of the series
analyzed here apart from the Netherlands series from 1301 and the
German series from AD 1000 (in Supplement~A [\citet{KOG14a}]). However, it
is more likely that the Little Ice Age represents recurring episodes
of worsened climate that should be detectable here.}

Our concern in this paper is not with the Little Ice Age globally
or across the Northern Hemisphere, but its extent in Europe. A central
lesson from efforts to reconstruct global temperature over the last
15 years is that regional climates vary widely and are therefore noisy
signals of hemispheric climate, so that the findings that we present
for Europe have no necessary implications for the behavior of average
temperature over the wider Northern Hemisphere.

A combination of resonant images invoked by \citet{hLam95} has linked
the Little Ice Age firmly to Northern Europe. These include the collapse
of Greenland's Viking colony and the end of grape-growing
in southern England in the fourteenth century; the Dutch winter landscape
paintings of Pieter Bruegel \mbox{(1525--1569)} and Hendrik Avercamp (1585--1634);
the periodic ``ice fairs''
on London's
Thames, ending in 1814; and, as the Little Ice Age waned, the contraction
of Europe's Nordic and Alpine glaciers. Among historians,
the idea that major economic and political events during the Little
Ice Age, and particularly during the seventeenth century, were driven
by worsening climate has generally been treated with scepticism [see,
e.g., \citet{aApp81} and \citet{jDV81}], but a notable
exception is \citet{gPar08} who attributes most of the political
turbulence of the \mbox{mid-seventeenth} century, from the Irish Catholic
uprising to the Mughal civil war, to a supposed worsening of climate
during this period.

\subsection{Data sources}

In this paper we analyze weather reconstructions for Europe that are
based on documentary sources. An immediate question is why more systematic
proxies such as tree rings cannot be used instead. The answer is that
tree rings accurately reflect annual weather conditions only in places
where trees are at the edge of their geographical range, under stress
because of cold or aridity; something that is not the case in most
of Europe outside Scandinavia.\footnote{More controversially, \citet{McSWei11} argue that
currently used
proxies such as tree rings, lake sediments and ice cores have low
explanatory power for recorded Northern Hemisphere temperature since
1850.}

Instead, for the period before 1700, when instrumental records begin,
an abundance of material allows European temperatures to be reconstructed
from documentary sources. For the sixteenth and seventeenth centuries,
weather diaries and ships logs exist in considerable numbers. For
earlier centuries, information about weather conditions is available
from recorded harvest dates for grains, hay and grapes, and, in particular,
from records of river tolls and water mills: how long each year were
rivers unnavigable or mills unusable because waterways were frozen
in winter or dried up in summer. A useful survey of these documentary
sources is given by \citet{BPW05}.

The two pioneering documentary reconstructions of regional European
weather are monthly Swiss temperature and precipitation from 1525
[\citet{cPfi92}] and quarterly Netherlands temperature from 1301
[\citet{vanE01}],\footnote{Although these series start in AD 800, there are increasing numbers
of missing observations as we go back past 1301 and the authors are
less confident of their accuracy, putting them in wider bands that
they denote by Roman rather than Arabic numerals. In running times
series tests, missing observations during the 14th and early 15th
centuries (30 for winter, 11 for summer) were set at the median value
of the entire series.} and we analyze both here. The current definitive reconstruction is
the monthly Central Europe reconstruction since 1500 by \citet
{Dobrovolny2009},
which includes authors of most of the previous major European weather
reconstructions and attempts to improve calibration of documentary
records against instrumental records by trying to correct instrumental
records for urban heat island effects and the impact of switching
the location of thermometers from north-facing walls to modern louvered
boxes. The monthly Central England series of \citet{gMan74} is based
entirely on instrumental records, albeit with some heroic data splicing
before 1700, and is included to look at weather in a more oceanic
zone of Europe.

Documentary estimates of German temperature have been extended back
to 1000 AD by \citet{GlaRie09} who label years as good, average
or bad. Because these data are multinomial we analyze them separately
in Supplement~A [\citet{KOG14a}]. It is worth noting that although our early
data are documentary, data for the eighteenth and early nineteenth
centuries, which lie within most definitions of the Little Ice Age,
are instrumental records.

\subsection{Reliability of documentary reconstructions}

How believable are these documentary reconstructions? With the exception
of the Netherlands series, which provides detailed accounts of its
construction (in Dutch), most studies give little detailed information
on sources and methodologies used to translate documentary records
into temperatures. However, we can still validate these series by
seeing how they correlate with records of agricultural activity not
used in their reconstruction.

\begin{figure}

\includegraphics{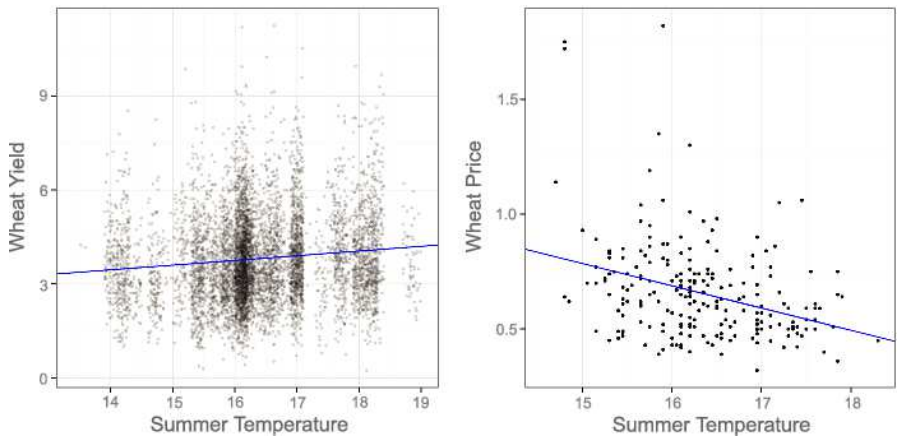}

\caption{\textup{(Left)} wheat yield per seed on 144 English manors versus
Netherlands summer temperature, 1270--1450.
\textup{(Right)} log English wheat price versus average Netherlands temperature over previous two years, 1213--1500.}\label{figWheat}
\end{figure}

The most detailed and extensive records of agricultural
output in Europe before the establishment of research stations at
the end of the nineteenth century are the accounts kept by English
manors between the thirteenth and fifteenth centuries, which have
been tabulated by \citet{Cam07}. The left-hand panel of Figure~\ref{figWheat}
plots Netherlands summer temperatures against the annual ratio of
wheat harvested to wheat sown on 144 manors from the start of accurate
records in 1270 (the earliest accounts start in 1211, but some of these
early records claim anomalously high yields that are not reflected
in low wheat prices: including these records did not affect the results
materially) and end in 1450 by when this pattern of seigneurial agriculture
carried out by coerced labor had virtually disappeared. Points are
jittered to separate overlapping ones.

%

It can be seen that yields move roughly in line with the Netherlands
temperature estimates of \citet{vanE01}, although explanatory power
in Table~\ref{TWheat} is low: a one degree rise in summer temperature
increases the average yield ratio by 5 per cent, while winter temperature
has no impact. Estimating the regression with mixed effects to allow
coefficients to vary across manors did not reveal large variation
across manors or change the reported estimates markedly. As a further
test of the validity of these results, we looked at the impact of temperature
on the yields of barley and oats, which are known to be more weather
resistant than wheat, and found that summer temperature had a smaller
effect on barley and none on oats.

English medieval agriculture was highly commercialized,
and price series for wheat exist back to 1211 [\citet{gCla04}].\footnote{We move Clark's price series back by one year to align them
with calendar
years rather than harvest years.} Wheat could be stored for a year after harvesting, so prices reflect
the previous two years' harvests: one poor harvest had a limited impact,
but successive harvest failures (such as occurred during the Great
Famine from 1316--1317, when wheat prices rose to nearly three times
their average level---shown by the two points in the northwest corner
of the second panel of Figure~\ref{figWheat}) were lethal. We show
elsewhere [\citet{mKcO10}] that death rates at all levels, from unfree
tenants to the high nobility, rose sharply after poor harvests, which
caused epidemic disease to spread across society.

%
\begin{table}
\tabcolsep=0pt
\caption{Regression of annual English wheat prices (1211--1450) and
yields (1270--1450) on estimated Netherlands temperatures}\label{TWheat}
\begin{tabular*}{\tablewidth}{@{\extracolsep{\fill}}@{}ld{2.4}d{2.4}d{2.4}d{2.4}d{2.4}ccd{4.0}@{}}
\hline
&\multicolumn{1}{c}{\textbf{Intercept}} & \multicolumn{1}{c}{\textbf{Summer}} & \multicolumn{1}{c}{\textbf{Lag summer}}
&\multicolumn{1}{c}{\textbf{Winter}} & \multicolumn{1}{c}{\textbf{Lag winter}} &\multicolumn{1}{c}{\textbf{RMSE}}
&\multicolumn{1}{c}{$\bolds{R^2}$} & \multicolumn{1}{c@{}}{$\bolds{N}$}\\
\hline
Yield & 0.504 & 0.046 & & 0.001 & & 0.369 & 0.016 & 6037 \\
& (0.077) & (0.005) & & (0.003) & & & & \\
Price & 5.771 & -0.050 & -0.048 & -0.003 & -0.010 & 0.261 & 0.076 & 164\\
& (0.454) & (0.019) & (0.020) & (0.012) & (0.012) & & & \\
\hline
\end{tabular*}
\tabnotetext[]{tt1}{Regression of annual English wheat prices and yields (both in logs) on current and lagged
Netherlands summer and winter temperature. Standard errors in parentheses.}
\end{table}

We analyze wheat prices from 1211 until 1500, a period during which
the general price level was stable before the Price Revolution of
the sixteenth century. Regressing log price on current and lagged
summer and winter temperatures in Table~\ref{TWheat}, it can be
seen that a one degree rise in summer temperature reduced prices by
5 per cent in the current and following year, while, again as we saw
with yields, winter temperature has no discernible impact. In summary,
then, its ability to predict medieval English wheat yields and prices
suggests that the Netherlands temperature reconstruction is a reliable
one.

\section{Change points in temperature since the Middle Ages}\label{secChange-Points-in}

To examine how weather deteriorated during the Little Ice Age, we analyze
several widely used annual summer and winter temperature reconstructions
up to 2000 for Western Europe: Central Europe from 1500; Netherlands
from 1301; Switzerland from 1525; and England from 1660.\footnote{We look at the \citet{LDX04} reconstruction of average
temperature
across all of Europe since 1500 and the \citet{GlaRie09} reconstruction
of German temperatures since AD 1000 in Supplement~A [\citet{KOG14a}].} The Central Europe series is expressed as a deviation in degrees
from the 1961 to 1990 average; monthly Swiss temperatures are assigned
to an integer scale from plus three (very good) to minus three (very
bad), while the Netherlands and English series are expressed in degrees
Celsius. We subtract the mean of the Netherlands and English series
prior to analysis, and look at summer (defined as the mean from June
to August) and winter (defined as the mean from December to February)
temperatures.

In this section we look for change points in
average temperature series, focusing on the classical change point
tests of \citet{jBpP98} and \citet{eVaO07} and the Bayesian test
of \citet{BarryHartigan1993}. In climatological terms, changes
in forcing variables are likely to be associated with long trends
rather than changes in levels, and we look explicitly at tests for
temporal dependence in annual temperature in the next section. However,
changing means associated with varying time trends are readily detectable
with change point analyses used in this section.

To look at mean variations, we start with a simple one-way ANOVA to
examine winter and summer temperature by a half century for each series,
and then compare variance of temperature within decades with variance
between decades, before moving on to change point tests.

\subsection{One-way ANOVA}

Temperature in year $i$ during half-century $j$ is assumed to be
normally distributed $y{}_{ij}\sim N(\alpha_{j},\sigma_{w}^{2})$,
while for mean temperature during half-century $j$ $\alpha_{j}\sim N(\mu,\sigma_{b}^{2})$.
To identify coefficients, each series is demeaned, and we constrain
the mean temperatures over half centuries to sum to zero $\Sigma
_{j}\alpha_{j}=0$,
and impose standard noninformative\vspace*{1pt} priors: $\mu\sim N$(0, 10,000),
$\sigma_{b}\sim U[0,20]$, $\sigma_{w}\sim U[0,20]$. This was estimated
by MCMC in JAGS with 10,000 iterations, the first 2500 being
discarded. Trace plots indicate rapid convergence on the posterior
distribution, and the Gelman--Rubin diagnostic supports convergence.

%
\begin{figure}

\includegraphics{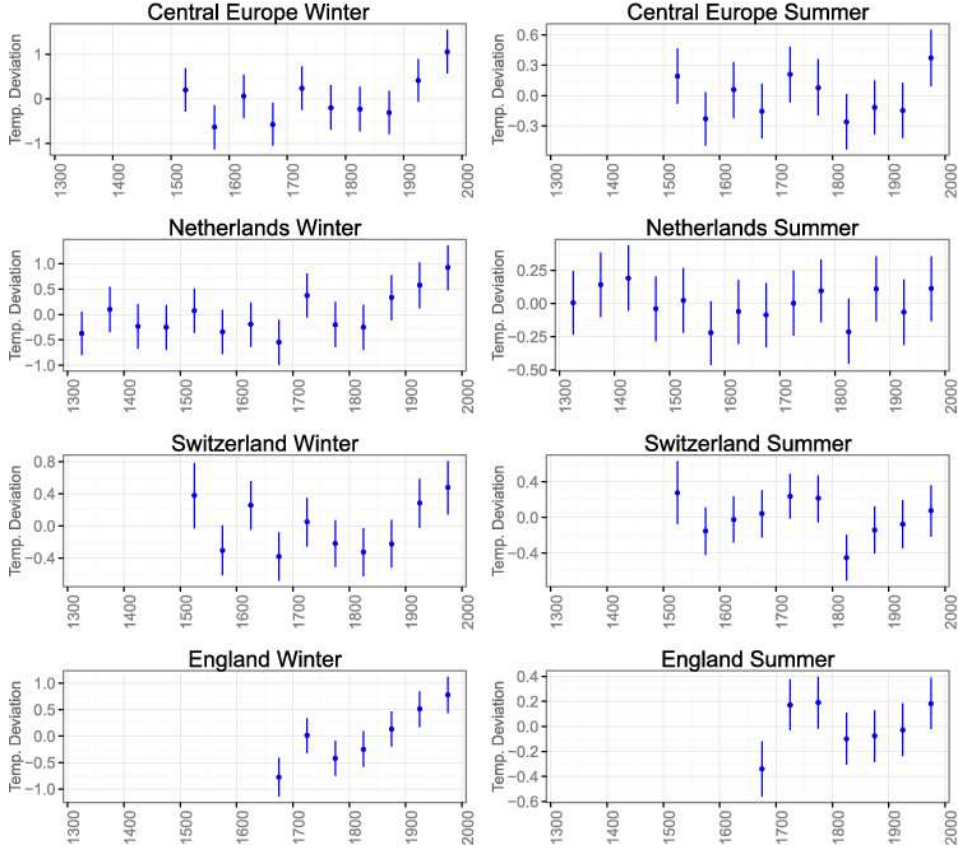}

\caption{Mean summer and winter temperature deviation by half century
with 95 per cent credible intervals.}\vspace*{-1pt}\label{figanova}
\end{figure}

Figure~\ref{figanova} shows little variation in summer temperature
with most observations lying within 0.25 degrees C of the series mean.
For winter series and Central European summers the rise around the
late nineteenth century is evident.

\subsection{Intra-class correlation}

Given that variance appears fairly constant for
most series (see Supplement~A [\citet{KOG14a}]), it is worthwhile to compare the
variance of temperature between periods with the variance within periods.
We use the same specification, priors and number of iterations for
annual temperature as in Figure~\ref{figanova}, but use decades
rather than half centuries as our period of analysis (the results
are almost identical if we use half centuries), and end each series
in 1900.

%
\begin{table}
\tabcolsep=0pt
\caption{Within and between decade standard deviation of annual temperature series before 1900}\label{TIcc}
\begin{tabular*}{\tablewidth}{@{\extracolsep{\fill}}@{}lcccccc@{}}
\hline
&\multicolumn{3}{c}{\textbf{Winter}} & \multicolumn{3}{c@{}}{\textbf{Summer}}\\[-6pt]
&\multicolumn{3}{c}{\hrulefill} & \multicolumn{3}{c@{}}{\hrulefill}
\\
&\multicolumn{1}{c}{\textbf{ICC}} & \multicolumn{1}{c}{$\bolds{\sigma_w}$} & \multicolumn{1}{c}{$\bolds{\sigma_b}$}
&\multicolumn{1}{c}{\textbf{ICC}} & \multicolumn{1}{c}{$\bolds{\sigma_w}$} & \multicolumn{1}{c@{}}{$\bolds{\sigma_b}$}\\
\hline
C. Europe & 0.01 & 1.92 & 0.11 & 0.04 & 1.07 & 0.19 \\
& [0, 0.04] & [1.79, 2.06] & [0, 0.36] & [0, 0.1] & [1, 1.15] & [0.03, 0.36]
\\[2pt]
Netherlands & 0.02 & 1.64 & 0.18 & 0.04 & 0.89 & 0.17 \\
& [0, 0.06] & [1.55, 1.74] & [0.01, 0.4] & [0, 0.1] & [0.84, 0.95] & [0.04, 0.29]
\\[2pt]
Switzerland & 0.03 & 1.15 & 0.16 & 0.05 & 0.97 & 0.2 \\
& [0, 0.09] & [1.07, 1.23] & [0.02, 0.36] & [0, 0.13] & [0.9, 1.05] & [0.05, 0.37]
\\[2pt]
England & 0.04 & 1.34 & 0.23 & 0.07 & 0.78 & 0.21 \\
& [0, 0.14] & [1.23, 1.47] & [0.01, 0.52] & [0, 0.2] & [0.71, 0.86] & [0.05, 0.38] \\
\hline
\end{tabular*}
\tabnotetext[]{tt2}{Intraclass correlation,
between decade standard deviation and within decade standard deviation
of annual temperature series. 95 per cent credible intervals in brackets.}\vspace*{-4pt}
\end{table}

Table~\ref{TIcc} reports the estimated standard deviation within
decades $\sigma_{w}$ and between decades $\sigma_{b}$, and also
the intra-class correlation, the\vspace*{1pt} percentage of the variance of the
series accounted for by between-class variance: ICC$\equiv\sigma
_{b}^{2}/ (\sigma_{b}^{2}+\sigma_{w}^{2} )$.
It can be seen that the standard deviation of temperature between
decades is low: of the order of one quarter of a degree Celsius. Similarly,
the intra-class correlation is low, with between-decade variance accounting
for one to three per cent of variance for winter temperature, and
four to eight per cent for summer.

\subsection{Classical change point tests}

We now look at the stability of each series: can we find structural
breaks in mean temperature corresponding to different phases of climate?
We start with classical tests: the \citet{BP03} procedure, implemented
by \citet{ZLH02}, which uses linear programming to find the optimal
location of $k$ breakpoints that minimize the sum of squared residuals,
subject to a Bayes Information Criterion penalty for adding breakpoints;
and the \citet{eVaO07} circular modification of \citet
{aSmS75} binary
segmentation which looks for the largest change in the partial sums
of observations. The Bai--Perron test looks for instability in regression
coefficients but, because we find little evidence for large or significant
trends or autocorrelation in the weather series (see Section~\ref{subFirst-Order-Autoregressions}
below), we focus on shifts in the intercept or mean value here.\footnote{The older CUSUM test [implemented by \citet{ZLH02}]
performed poorly
in simulations, only detecting half as many breaks in short series
as Bai--Perron, and we do not report its results here. Because we are
using seasonal averages, and only have annual data, the \citet{gDjT}
test for changes in the parameter of the Pareto distribution generating
extreme values is not applicable.} Climatology would lead us to expect shifts in temperature trends
rather than in levels, but in Section~\ref{subFirst-Order-Autoregressions}
we find no indication of substantial trends or structural breaks in
trends before 1900.

For these tests to be informative, we must know their power: are they
capable of detecting shifts in mean temperature of the sort that might
have occurred during a European Little Ice Age? We will examine the
power of these tests to find changes of means in series which are
independent draws from a normal distribution: we will see in
Section~\ref{secConditional-Independence} that our data are compatible
with the assumption of no temporal dependence.

Starting with the case of a series of length 300 with a constant mean
of zero (in all simulations the standard deviation is unity), in 1000
simulations, BP detected breaks in 1.8 per cent of cases and VO in
1.4 per cent, close to its nominal significance level of 1 per cent.
In all cases here, the minimum segment length in BP is set to 15 per
cent of the data series. Looking at the ability to detect changes,
in 1000 simulations where 150 observations of mean zero are followed
by 150 with a mean of 0.5, BP detected the break in 90 per cent of
cases and VO in 64 per cent, while for 100 observations in each group
the success rates are 71 per cent and 43 per cent. By contrast, the
Bayesian change point analysis of \citet{BarryHartigan1993}, which
performs better than classical tests in detecting short breaks, finds
only around one quarter of 0.5 standard deviation changes halfway
through a series of length 200 or 300.\footnote{For a change in the posterior mean to be detectable by eye,
it requires
a posterior probability of a change point of at least 0.15, which,
looking at the maximum posterior probability for 10 observations on
either side of the break, occurs in around 25 per cent of simulations.}

Looking at a series of 150 observations where the middle 50 are 1
standard deviation higher, BP detects 96 per cent and VO 95; for a
rise of 0.75 the percentages detected are 71 and 65, while for a rise
of 0.5 the detection rates were 28 and 22 per cent. For a rise in
33 observations in the middle of a series of 100, for a one standard
deviation increase BP detected 84 per cent of cases and Segment~77;
for a rise of 0.75 the rates are 52 and 40, while for a rise of 0.5
the detection rates are 21 and 13. In other words, binary segmentation
and BP can detect changes of~1 standard deviation in annual temperature
(roughly 1 degree Celsius for summer temperatures in Northern Europe)
that last a generation and fairly reliably detect 0.5 standard deviation
changes that last a century.

%
\begin{table}
\tabcolsep=0pt
\caption{Change points in mean winter and summer temperatures}\label{TBreaks}
\begin{tabular*}{\tablewidth}{@{\extracolsep{\fill}}@{}lccccc@{}}
\hline
&&\multicolumn{2}{c}{\textbf{Winter}} & \multicolumn{2}{c@{}}{\textbf{Summer}}\\[-6pt]
&&\multicolumn{2}{c}{\hrulefill} & \multicolumn{2}{c@{}}{\hrulefill}
\\
&\multicolumn{1}{c}{\textbf{Start}} & \multicolumn{1}{c}{\textbf{BP}} & \multicolumn{1}{c}{\textbf{VO}}
&\multicolumn{1}{c}{\textbf{BP}} & \multicolumn{1}{c@{}}{\textbf{VO}}\\
\hline
C. Europe&1501&1909&1909&0&1982\\
Netherlands&1301&1861&1897&0&0\\
Switzerland&1525&1910&1911&0&1813, 1818\\
England&1660&1910&1911&0&0\\
\hline
\end{tabular*}
\tabnotetext[]{tt3}{Change points in mean winter and summer temperatures identified by Bai--Perron breakpoints and
Venkatraman--Olshen binary segmentation.}
\end{table}

Table~\ref{TBreaks} reports the change points
detected in our four series of summer and winter temperature using
BP and VO. In all cases we find a break in winter temperatures around
the start of the twentieth century, but no indication of any change
\mbox{before} that. For summer temperature, England and the Netherlands show
no change points. Central Europe shows a rise in the late twentieth
century, while Switzerland records a shift between 1813 and 1818;
we return to this below.

In summary, classical change point tests suggest that if sustained
falls in temperature did occur in Europe during the Little Ice Age,
their magnitude was below half a standard deviation.

\subsection{Bayesian change points}

%

\begin{figure}

\includegraphics{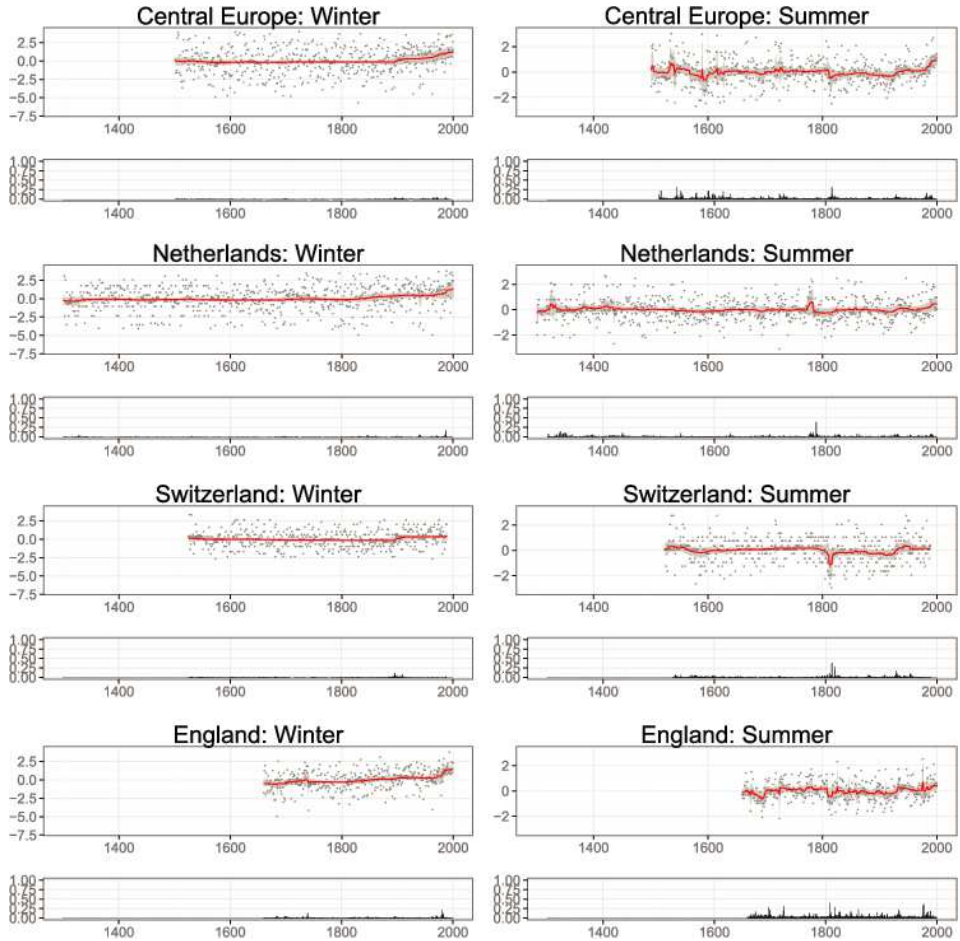}

\caption{Posterior mean and probability of breakpoint for temperature
deviation series using the Barry and Hartigan product partition model.}\label{figBreaks}
\end{figure}

The breakpoint and segmentation methods perform
well in detecting long-lasting changes in series, but do less well
at finding shorter breaks. We therefore consider the Bayesian change
point analysis of \citet{BarryHartigan1993} implemented through
the MCMC approximation of \citet{cEjW07}. Figure~\ref{figBreaks}
shows the estimated mean of each series inside a 95 per cent credible
interval, with the posterior probability of a change point plotted
below, all estimates being carried out using the default values of
\citet{cEjW07}.

It is evident that winter temperatures are stable until the twentieth
century when they rise markedly, particularly for England. While designed
to detect changes in means, upward trends associated with post-industrial
warming in the twentieth century are readily apparent. For summers,
the older reconstructions for Netherlands and Swiss temperature show
little variation before the twentieth century, apart from a rise in
the late eighteenth century for the Netherlands, and a drop in the
1810s for Switzerland.

%

The English and the Central European series, while not showing sustained
changes before 1900, do show considerable volatility around their
mean, with episodes of sustained falls in summer temperature lasting
around a decade. The most notable of these, that also appears in many
of the city series included in the \hyperref[append]{Appendix}, occurs in the 1810s when
temperatures in England were below average every year between 1809
and 1817, and in Central Europe between 1812 and 1818. Similarly,
temperatures in Central Europe were below average every year from
1591 to 1598, and in England from 1687 to 1698.\footnote{It is notable that some of these falls occur in the same
decade as
well-known episodes of volcanism such as Tambora in 1815, and Hekla,
Amboina and Serua in 1693--1694, although in both cases the temperature
falls precede the volcanic episodes by several years.} While summer temperatures do not show the prolonged changes that
one would expect during a Little Ice Age, there are occasional decades
of notably worse weather.

\section{Temporal dependence in temperature}
\label{secConditional-Independence}

While European temperatures do not show much sign of discrete changes,
Little Ice Age events are more likely to have been associated with
gradual changes.\footnote{In addition, as a referee observes, change points may be eliminated
in documentary reconstructions if their authors splice data together
by assigning early reconstructions the same mean as later instrumental
observations. Such homogenization will not eliminate the temporal
dependence that we test for in this section.} We therefore examine the temperature reconstructions for temporal
dependence. We start with a simple first order autoregression with
trend, and then consider tests of conditional mean independence that
are sensitive to a range of higher order and nonlinear dependencies.

\subsection{First order autoregressions}\label{subFirst-Order-Autoregressions}

%
\begin{table}
\tabcolsep=0pt
\caption{Regression of annual temperature reconstructions on lagged temperature and trend}\label{TLinear}
\begin{tabular*}{\tablewidth}{@{\extracolsep{\fill}}@{}lcd{2.4}d{2.4}d{2.4}cc@{}}
\hline
& \multicolumn{1}{c}{\textbf{Start}} & \multicolumn{1}{c}{\textbf{Intercept}}
& \multicolumn{1}{c}{\textbf{Lag temperature}} & \multicolumn{1}{c}{\textbf{Trend}}
& \multicolumn{1}{c}{\textbf{RMSE}} & \multicolumn{1}{c}{$\bolds{R^2}$}\\
\hline
\multicolumn{7}{@{}c@{}}{\textit{Summer}}\\
C. Europe& 1501 &-1.145 & -0.108 & 0.000 & 1.908 & 0.012\\
& & (0.199) & (0.050) & (0.001) & &
\\[3pt]
Netherlands&1302& -0.267 & -0.039 & 0.000 & 1.649 & 0.004\\
& & (0.136) & (0.041) & (0.000) & &
\\[3pt]
Switzerland & 1526 &-0.266 & -0.138 & -0.001 & 1.144 & 0.028\\
& & (0.119) & (0.051) & (0.001) & &
\\[3pt]
England&1661& -0.601 & -0.051 & 0.003 & 1.342 & 0.020\\
& & (0.179) & (0.065) & (0.001) & &
\\[6pt]
\multicolumn{7}{@{}c@{}}{\textit{Winter}}\\
C. Europe& 1501 & -0.083 & 0.112 & 0.000 & 1.077 & 0.016\\
& & (0.109) & (0.050) & (0.000) & &
\\[3pt]
Netherlands&1302& 0.053 & 0.100 & 0.000 & 0.904 & 0.012\\
& & (0.074) & (0.041) & (0.000) & &
\\[3pt]
Switzerland & 1526 &0.100 & 0.063 & -0.001 & 0.981 & 0.012\\
& & (0.102) & (0.052) & (0.000) & &
\\[3pt]
England&1661& -0.034 & 0.102 & 0.000 & 0.795 & 0.011\\
& & (0.104) & (0.065) & (0.001) & & \\
\hline
\end{tabular*}
\tabnotetext[]{tt4}{Regression of annual temperature on lagged temperature and a trend. Standard errors in parentheses. All series end in 1900.}
\end{table}

Table~\ref{TLinear} gives the results of a
first order autoregression with trend $y_{t}=\alpha+\beta
y_{t-1}+\gamma t$
for each weather series. We shall see below that this specification
is adequate: there is little indication of higher order dependencies
or nonlinearities in the data. Each series is ended in 1900 and a
\citet{BP03} procedure does not indicate a structural break in any
regression associated with changes between rising and falling trends
of temperature.

For every series the regression $R^{2}$ is below 0.03. The size of
autocorrelation is small in every case, and only a few series show
statistically significant autoregression at conventional levels. Central
European and Netherlands winters show significant correlation but
with a coefficient around $0.1$: a one degree rise in temperature
one year increases average temperature the next winter by an almost
imperceptible
tenth of a degree. The significance of the Netherlands series is caused
by reconstructions from the fourteenth century and disappears from
the fifteenth century onward, while significance for the Central
Europe series disappears after 1700. Similarly, the small negative
autocorrelation in the Central Europe and Swiss winter series disappears
when noninstrumental observations before 1700 are excluded.

\subsection{Conditional mean independence}

While these temperature reconstructions do not, with some exceptions
for early periods, display first order autoregression, there remains
the possibility that the series show some other form of dependence,
either higher order linear or nonlinear. To investigate this possibility,
we analyze our weather series for what statisticians call conditional
mean independence and financial econometricians, who developed these
tests to see if changes in asset prices are unpredictable, call martingale
differences. Specifically, for a stationary series $\{Y_{t}\}$, let
$I_{t}=\{Y_{t},Y_{t-1},\ldots\}$ denote the information set at time
$t$. Under martingale differences $E (Y_{t}|I_{t-1} )=\mu$
[\citet{EL09b}].

\subsubsection{Tests}

Martingale differences imply zero autocorrelations $\rho_{i}$ for order $i>0$.
This leads to the standard Ljung--Box Portmanteau test based on the
sum of the first $p$ squared autocorrelations $T\Sigma_{i=1}^{p}\tilde
{\rho_{i}}^{2}$,
where $\tilde{\rho_{i}}=\rho_{i}\sqrt{ (T+2 )/ (T-i )}$
and $T$ is the number of observations. We apply the modification
of \citet{Escanciano2009} where correlations are divided by sample
autocovariances of the squared series to provide robustness against
heteroskedasticity, and $p$ is chosen by a data-dependent procedure
where a penalty term is subtracted that switches between Akaike and
Bayes Information Criteria.

The second test we apply is a variance ratio test, based on the idea
of \citet{LMac89} that, for uncorrelated series, estimated variance
should rise in proportion to the length of the series: $\mathrm{AVR}=1+2\Sigma_{i=1}^{p-1} (1-i/p )\rho_{i}$.
We apply the \citet{Kim2009} modification where $p$ is chosen by
a data dependent procedure, and the distribution of the test statistic
is derived by applying a wild bootstrap, where each term $Y_{t}$
of the original series is multiplied by a random variable with zero
mean and unit variance.

The third class of tests for temporal dependence in time series are
tests for the departure of the series spectrum from linearity. In
econometrics these originate with \citet{Durlauf1991}, and we report
the generalized spectral test of \citet{Escanciano2006}.

\subsubsection{Results}
%
\begin{table}
\tabcolsep=0pt
\caption{Tests for conditional independence in means}\label{TMDS}
\begin{tabular*}{\tablewidth}{@{\extracolsep{\fill}}@{}lcccccc@{}}
\hline
&\multicolumn{3}{c}{\textbf{Summer}} & \multicolumn{3}{c@{}}{\textbf{Winter}}\\[-6pt]
&\multicolumn{3}{c}{\hrulefill} & \multicolumn{3}{c@{}}{\hrulefill}\\
&\multicolumn{1}{c}{$\bolds{Q}$} & \multicolumn{1}{c}{\textbf{VR}} & \multicolumn{1}{c}{\textbf{Spec}} & \multicolumn{1}{c}{$\bolds{Q}$}
&\multicolumn{1}{c}{\textbf{VR}} & \multicolumn{1}{c@{}}{\textbf{Spec}}\\
\hline
\multicolumn{7}{@{}c@{}}{1701--1900}\\
C. Europe&0.23&0.06&0.43&0.29&0.00&0.78\\
Netherlands&0.13&0.05&0.10&0.30&0.24&0.86\\
Switzerland&0.53&0.43&0.25&0.17&0.12&0.29\\
England&0.42&0.42&0.51&0.46&0.60&0.62
\\[3pt]
\multicolumn{7}{@{}c@{}}{1701--2000}\\
C. Europe& 0.01& 0.02 &0.01& 0.32 &0.32& 0.83\\
Netherlands & 0.05& 0.01& 0.10 &0.63 &0.64 &0.62\\
Switzerland &0.01& 0.26& 0.19& 0.71& 0.74& 0.47\\
England& 0.22& 0.14& 0.32& 0.00 &0.04& 0.04
\\[3pt]
\multicolumn{7}{@{}c@{}}{\textit{pre}-1701}\\
C. Europe&0.03&0.01&0.01&0.07&0.00&0.09\\
Netherlands&0.24&0.00&0.29&0.61&0.92&0.85\\
Switzerland&0.02&0.25&0.03&0.45&0.36&0.20\\
England&0.16&0.15&0.40&0.06&0.04&0.05
\\[3pt]
\multicolumn{7}{@{}c@{}}{\textit{pre}-1901}\\
C. Europe&0.03&0.02&0.02&0.04&0.00&0.23\\
Netherlands&0.01&0.00&0.00&0.38&0.41&0.58\\
Switzerland&0.19&0.13&0.54&0.01&0.65&0.01\\
England&0.00&0.03&0.09&0.62&0.79&0.48\\
\hline
\end{tabular*}
\tabnotetext[]{tt5}{$p$-values for tests of
conditional independence of means of temperature series until 2000. $Q$~is a~robustified portmanteau test with automatic lag selection.
VR gives the wild bootstrap test results for an automatic variance
ratio test. Spec is a generalized spectral test.}
\end{table}

Looking at the small sample properties of these
tests for samples with 100, 300 and 500 observations, \citet{Charles2011}
find that the reported size of all tests is approximately correct
and that against models of linear dependence, the automated variance
ratio test shows highest power, while against a variety of nonlinear
processes, the generalized spectrum test works best. The temperature
series here do not appear to exhibit nonlinearity: applying a
Teraesvirta--Lin--Granger
(\citeyear{TLG93}) test for
nonlinearity in means using the code
of \citet{tser10} led to $p$-values in excess of 0.09 for all series,
with values above 0.5 in most cases.

Table~\ref{TMDS} reports the $p$-values of the three temporal dependence
tests for each temperature series, calculated using the default values
of \citet{vrtest10}. The first block reports results for each series
from 1701 to 1900; the second is for all years after 1701; the third
has each series from its start (1500 for Central Europe, 1301 for
the Netherlands, 1525 for Switzerland and 1660 for England) until
1700; while the final block gives results for the pre-1901 period.

It can be seen that for 1701--1900 (a period whose start lies in conventional
definitions of the Little Ice Age) the only test that rejects conditional
mean independence at conventional levels is the VR test for European
winters. This seems to result from excessive sensitivity of the VR
test: this series has a first order autoregressive coefficient of
$-$0.06 with a $p$-value of 0.36; and if the residuals from this
first order autoregression are tested, the VR test returns a $p$-value
of 0.82 (with the automatic Portmanteau and generalized spectrum test
giving similar values), indicating that the first-order specification
is adequate. Adding in twentieth-century observations in the second
block, the only series to show systematic departures from conditional
mean independence are Central European summers and English winters,
both of which rise notably after 1900.

The excess sensitivity of the VR test also appears in the pre-1701
Central Europe winter data: the coefficient of a first order autoregression
is $-$0.14 with a $p$-value of 0.05, and applying the VR test to
residuals gives a $p$-value of 0.97. For the summer data, the coefficient
of a first order autoregression is $0.12$ with a $p$-value of 0.09,
and applying the VR test to residuals gives a $p$-value of 0.82.
We find similar behavior in pre-1701 Netherlands temperatures: the
coefficient of a first order autoregression is $-$0.04 with a $p$-value
of 0.38, and applying the VR test to residuals gives a $p$-value
of 0.87.

The weak correlation of annual Northern European temperature series
is in marked contrast to the strong autocorrelation in the CRU Northern
Hemisphere temperature series since 1850 which has first order autocorrelation
0.6 and significant partial autocorrelations out to lag four
[\citet{McSWei11}],
highlighting once again the importance of spatial variation in climatic
patterns.

\section{The Slutsky effect}
\label{secThe-Slutsky-Effect}

As we noted in the \hyperref[sec1]{Introduction}, our failure to find change points
or temporal dependence in European weather possibly reflects the fact
that we are analyzing unsmoothed data. By contrast, climatologists
tend to smooth data using a moving average or other filter prior to
displaying. When data are uncorrelated, as annual European weather
series appear to be, smoothing can introduce the appearance of irregular
oscillations: a Slutsky effect.

Confusingly, there are two different definitions of the Slutsky effect
in common use.\footnote{We are grateful to a referee for pointing this out.} First there is the formal sense, going back to \citet{eSlu37}, that
applying a filter to a white noise series will generate \emph{regular}
cycles corresponding to peaks in the transfer function of the filter.
For an $m$ period moving average, for example, the transfer function
is
\[
f (\omega)=\frac{1}{m^{2}} (1-\cos m\omega)/ (1-\cos\omega),
\]
which, for $m=25$, has its largest peak, after zero, around 17.5
years: too short, clearly, to generate Little Ice Age behavior.

The second sense, and the one that we invariably use here, is the
colloquial one that applying a moving average to a white noise series
will generate the appearance of \emph{irregular} oscillations,\vspace*{1pt} as
the filter is distorted by runs of high or low observations.\footnote{This is the definition given, for instance, at
\url{http://mathworld.wolfram.com/Slutzky-YuleEffect.html}.} The intuition behind the Slutsky effect is straightforward: just
as tossing a fair coin leads to long sequences with an excess of heads
or tails, so random sequences in general will occasionally throw up
some unusually high or low values in close succession that will distort
a smoothing filter. In climatology, \citeauthor{wBur03} [(\citeyear{wBur03}), page~24] briefly discusses
the Slutsky effect, in the second sense, in an early chapter on statistical
background and gives a diagram illustrating how applying a moving
average to a series of random numbers will give the appearance of
irregular cycles, but does not subsequently investigate whether it
can be the source of perceived climate cycles.

%
\begin{figure}

\includegraphics{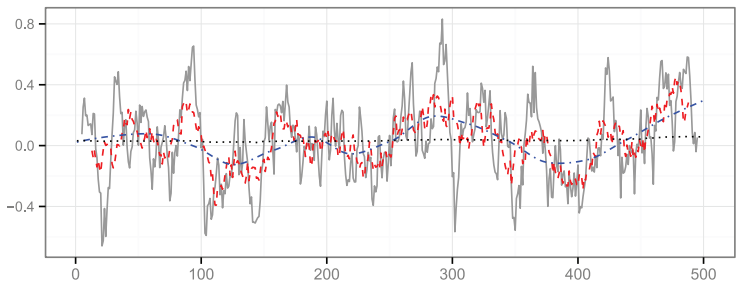}

\caption{Slutsky effect: 500 independent, standard normal random numbers smoothed
with a 10 period moving average (solid line);
25 period moving average (dashed line);
and loess smoother with span of one third (dotdash line).
The central dotted line gives the posterior mean estimated by a Barry--Hartigan change point procedure.}\label{figSlutsky-effect}
\end{figure}

The Slutsky effect is illustrated in Figure~\ref{figSlutsky-effect}
which gives smoothed values of~500 independent standard normal variables,
using moving\vspace*{1pt} averages of 10, 25 and 50, and R's loess filter with
smoothing span of one third.\footnote{The variables were generated in R with seed set to 123. The loess
smoother applies a polynomial regression of order two to points within
the smoothing span, with observations within the window being averaged
with tri-cubic weights. The loess smoother behaved almost identically
to the Butterworth low pass filter with threshold of 0.025, except
at the boundaries where the latter showed characteristic attenuation
toward zero.} It can be seen that a notable downward dip appears to occur after
observation 100, and particularly between observations 300 and 400,
which is followed by a marked upward trend. By contrast, the posterior
mean estimated by a \citet{BarryHartigan1993} change point procedure,
which we have seen to be particularly useful for detecting short changes
in the mean value of a series, shows little variation.

%
\begin{figure}

\includegraphics{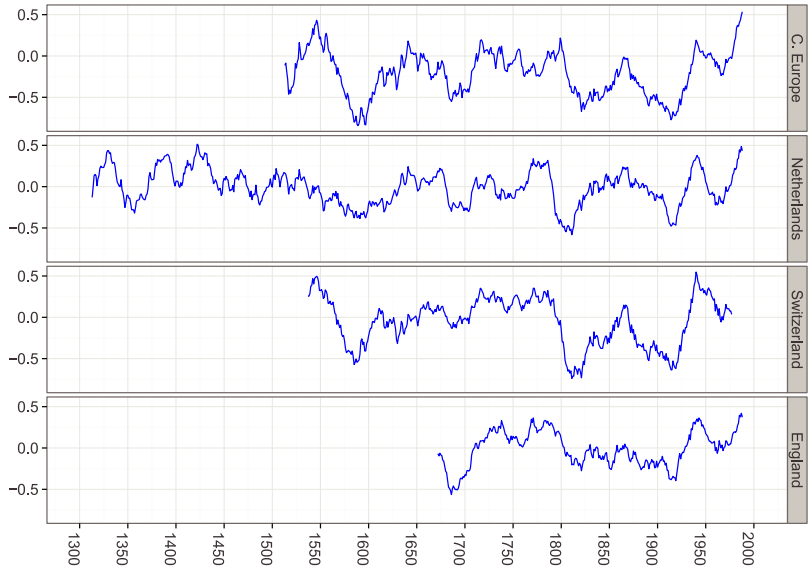}

\caption{Slutsky effect in European summer temperature deviation reconstructions,
1301--2000 smoothed with a 30 year moving average.}\label{figlowcountries}
\end{figure}

The Slutsky effect is illustrated in Figure~\ref{figlowcountries}
for the four temperature series, each smoothed with a 30 year moving
average. The longest, Netherlands series, appears to show a cooling
trend from the mid-fifteenth to the early nineteenth centuries, with
markedly cold episodes in the late sixteenth, late seventeenth and
early nineteenth centuries. Particularly interesting is the Central
Europe series in the top panel. The temperature pattern resembles
the \citeauthor{PAGES13} [(\citeyear{PAGES13}), Figure~2] European reconstruction (notably the
troughs in the late 1500s and 1600s, and the early 1800s) which is
based on the same series along with tree rings from the Pyrenees,
Alps, Balkans and Scandinavia.

Glaciers may be seen as a physical embodiment of a Slutsky effect:
their extent represents a moving average process of temperature and
precipitation over preceding years, and can show considerable variation
through time even though the annual weather processes that drive them
are independent draws from a fixed distribution. For example, while
annual Swiss winter temperature and precipitation are close to white
noise until the late nineteenth century (see Supplement~A [\citet{KOG14a}] for
precipitation results), Swiss glaciers fluctuate considerably, expanding
from the mid-fifteenth century until 1650, contracting until 1750
and then expanding again until 1850 [\citet{kBjM05}, pages~\mbox{18--19}].

\section{Conclusions}
\label{secConclusions}

Our intention was to estimate the extent and timing of climate changes
during the European Little Ice Age. To our surprise, despite multiple
tests, standard temperature reconstructions give little indication
of temporal dependence or sustained structural breaks in European
weather before the late\vadjust{\goodbreak} nineteenth century. If Europe experienced
a Little Ice Age, the weather reconstructions analyzed suggest that
temperature falls were of the order of less than half of one standard
deviation.

Our findings are a reminder that some of the changes claimed by
\citet{hLam95}
to be consequences of the Little Ice Age may have other possible causes.
The freezing of the Thames---which for most people is the most salient
fact about the Little Ice Age---was due to Old London Bridge which
effectively acted as a dam, creating a large pool of still water which
froze twelve times between 1660 and 1815 [\citet{jSch}, page 70]. Tidal
stretches of the river have not frozen since the bridge was replaced
in 1831, even during 1963 which is the third coldest winter (after
1684 and 1740) in the Central England temperature series that starts
in 1660.

For Greenland's Vikings, competition for resources
with the indigenous Inuit, the decline of Norwegian trade in the face
of an increasingly powerful German Hanseatic League, the greater availability
of African ivory as a cheaper substitute for walrus ivory, overgrazing,
plague and marauding pirates probably all played some role in their
demise [\citet{Bro00}]; and even if weather did worsen, the more
fundamental
question remains of why Greenland society failed to adapt [\citet{McG81}].
The disappearance of England's few vineyards is associated
with increasing wine imports after Bordeaux passed to the English
crown in 1152, suggesting that comparative advantage may have played
a larger role than climate.

Similarly, the decline of wheat and rye cultivation in Norway from
the thirteenth century may owe more to lower German cereal prices
than temperature change [\citet{hMis75}, page 59]. Moreover, with worsening
climate we would expect wheat yields to fall relative to the more
weather-robust spring grains barley and oats, whereas \citeauthor{ABC08} [(\citeyear{ABC08}), Table 1(A)~and~(B)],
find that between the early fifteenth and late seventeenth century,
wheat yields show no trend relative to oats and rise steadily relative
to barley.

Finally, demography supports our reservations about a European Little
Ice Age. We would expect Northern Europe to have shown weak population
growth as the Little Ice Age forced back the margin of cultivation.
In fact, while the population of Europe in 1820 was roughly 2.4 times
what it had been in 1500, in Norway the population was about 3.2 times
as large, in Switzerland 3.5 times, in Finland 3.9 times and in Sweden
4.7 times as large as in 1500 [\citet{aMad09}].

In summary, this paper makes two points: one methodological, one historical.
First, smoothing white noise or near white noise data is problematic,
but the most reliable and informative results, both in terms of avoiding
spurious oscillations and detecting real breaks, are given by the
\citet{BarryHartigan1993} procedure. Second, although most people
have strong priors that Europe experienced bouts of markedly worse
weather during the Little Ice Age, such episodes are not apparent
in standard temperature reconstructions.

\begin{appendix}\label{append}
\section*{Appendix: Data sources}
\begin{itemize}
\item Central European temperature reconstructions from 1500 by
\citet{Dobrovolny2009}
are available at \url{ftp://ftp.ncdc.noaa.gov/pub/data/paleo/historical/europe/dobrovolny2010temperature.xls}.
\item The Netherlands temperature series of \citet{vanE01} are available
at \url{http://www.knmi.nl/klimatologie/daggegevens/antieke_wrn/nederland\_wi\_zo.zip}.
\item Swiss summer and winter temperature and precipitation from\break 
\citet{cPfi92}
are available at \url{ftp://ftp.ncdc.noaa.gov/pub/data/paleo/historical/switzerland/clinddef.txt}.
\item Monthly mean Central England temperature from 1659 are from \url{http://hadobs.metoffice.com/hadcet/cetml1659on.dat},
and the monthly England and Wales precipitation series from 1766 are
from \url{http://hadobs.metoffice.com/hadukp/data/monthly/HadEWP\_monthly_qc.txt}.
\end{itemize}
\end{appendix}

\section*{Acknowledgments}
We would like to thank the Editor Tilmann Gneiting, an Associate Editor
and anonymous referees for their detailed and constructive
criticisms of earlier versions of this paper. All errors and
interpretations are ours. This research was undertaken as part of the HI-POD
(Historical Patterns of Development and Underdevelopment: Origins and
Persistence of the Great Divergence) Project supported by the
European Commission's 7th Framework Programme for Research.

\begin{supplement}
\sname{Supplement A}\label{suppA}
\stitle{Additional weather series\\}
\slink[doi]{10.1214/14-AOAS753SUPPA} 
\sdatatype{.pdf}
\sfilename{aoas753\_suppa.pdf}
\sdescription{This supplement analyzes additional weather series:
European cities since 1500; European average temperature since 1500;
German \mbox{temperature} since AD 1000;
and English and Swiss precipitation. It also examines the variance of
the temperature series examined here.}
\end{supplement}

\begin{supplement}
\sname{Supplement B}\label{suppB}
\stitle{Data and code\\}
\slink[doi]{10.1214/14-AOAS753SUPPB} 
\sdatatype{.zip}
\sfilename{aoas753\_suppb.zip}
\sdescription{This file contains the data and R code used in the paper.}
\end{supplement}


%

\printaddresses
\end{document}